\documentclass{elsart}

\usepackage{epsfig}

\usepackage{amssymb}

\newcommand{\be}{\begin{equation}}
\newcommand{\ee}{\end{equation}}
\newcommand{\ba}{\begin{array}}
\newcommand{\ea}{\end{array}}
\newcommand{\baa}{\begin{array}}
\newcommand{\eaa}{\end{array}}
\newcommand{\bea}{\begin{eqnarray}}
\newcommand{\eea}{\end{eqnarray}}

\newcommand{\Dslash}{\not \! \! D}

\newcommand{\hc}{{\rm h.c.}}

\begin{document}

%\begin{flushright}
%IFT-UAM/CSIC-08-30 \\
%FTUAM-08-09
%\end{flushright}
\begin{frontmatter}
\begin{flushright}
IFT-UAM/CSIC-08-30 \\
FTUAM-08-09
\end{flushright}



\title{Gluino zero-modes for  calorons at finite temperature}


\author[label1]{Margarita Garc\'{\i}a P\'erez}
 \ead{margarita.garcia@uam.es}
\author[label1,label2]{Antonio Gonz\'alez-Arroyo}
 \ead{antonio.gonzalez-arroyo@uam.es}
\author[label1,label2]{Alfonso Sastre}
 \ead{alfonso.sastre@uam.es}
\address[label1]{Instituto de F\'{\i}sica Te\'orica UAM/CSIC}
\address[label2]{Departamento de F\'{\i}sica Te\'orica \\
       Universidad Aut\'onoma de Madrid, E-28049--Madrid, Spain \\}

\begin{abstract}
We study the solutions of the Dirac equation  in the adjoint
representation(gluinos)  in the background field of  SU(2) unit charge  calorons.
Our solutions are forced to be antiperiodic in thermal time and would
occur naturally in a semiclassical approach to ${\cal N}=1$ Super-symmetric
Yang-Mills theory at finite temperature.
\end{abstract}

\begin{keyword}
Caloron \sep gluino zero-modes \sep finite temperature \sep super-symmetric
Yang-Mills theory 
%\PACS 
\end{keyword}
\end{frontmatter}

% main text
\section{Introduction}
\label{s.intro}
In this paper we  derive analytic expressions for the finite temperature gluino
zero modes of the Dirac operator in the background field of
the $Q=1$ SU(2) calorons. These are self-dual configuration in $R^3\times S_1$
including the well known Harrington-Shepard (HS) solution~\cite{HS} as well as the
non-trivial holonomy calorons~\cite{vanbaal0}-\cite{lee}.
The periodicity in one direction, to be referred as thermal-time, occurs
naturally in a path-integral approach to finite temperature Yang-Mills theory
and, with the inclusion of spinor fields in the adjoint representation (gluinos),
in its minimal supersymmetric extension. Calorons are thus the natural objects
to be considered in a semiclassical approach to these theories at finite
temperature.  They smoothly interpolate
between instantons and BPS monopoles at zero and high temperature
respectively~\cite{vanbaal0}-\cite{Rossi}, providing a very interesting
link between them.
One of the required ingredients for such semiclassical analysis is the
knowledge of fermionic zero modes in the background of the caloron field.
Although those in the fundamental representation of the gauge group
have been known for quite some time \cite{grossman}-\cite{maxim},
this is not the case for the gluino zero modes. They have been derived
only recently by two of the present authors~\cite{gluinoA}, and just for the
case of periodic boundary conditions in $S_1$. These are the relevant
modes for supersymmetric compactifications but not what
is needed when studying ${\cal N}=1$ SUSY Yang-Mills fields at finite
temperature. Antiperiodicity in thermal-time has to be required in that
case. The goal of this letter is to obtain analytic expressions for
the antiperiodic solutions, derived here for the first time
even for the trivial holonomy, Harrington-Shepard, case.
This requires a different approach than the one employed in~\cite{gluinoA}
which was based on the relation between zero modes and self-dual deformations
of the gauge field, providing only periodic solutions.

The paper is organized as follows. In section~\ref{s.form}
we will describe the strategy followed to obtain the antiperiodic
zero modes and present the analytic expressions for the solutions. In
section~\ref{s.prop} we analyze their properties
in several relevant limits, paying particular attention to the one
in which the caloron dissociates into a pair of static BPS constituent monopoles.
The trivial-holonomy HS zero mode solution and the equal mass constituent monopole
cases are also discussed in some detail.
Conclusions  and a brief summary of results are presented in section~\ref{s.concl}.

\section{Formalism}
\label{s.form}
As mentioned previously, our goal is that of solving  the massless covariant
Dirac equation in the adjoint representation of the group
\be
\label{D_Eq}
\Dslash \Psi=0
\ee
in the background field of a  Q=1 caloron~\cite{vanbaal0}-\cite{lee}.
This problem has been partially addressed in Ref.~\cite{gluinoA}.
The approach that was followed in that paper was based on the well-known
relation between self-dual deformations of the gauge field and the
zero-modes of the Dirac operator in the adjoint representation.
However, the solutions obtained in this way are periodic in
thermal-time with the same period $\beta$ (to be taken equal to 1 in
what follows) as the gauge field itself. Thus, a different strategy has
to be set up to derive the antiperiodic modes relevant for finite temperature.
In what follows we will present the basic idea behind
our procedure and the results obtained with it. In all technical
aspects we will rely strongly in  the notation and derivations done in
Ref.~\cite{gluinoA}.

The observation that leads to our solution is the fact that
antiperiodic solutions turn out to be periodic in the double period.
Thus, the method of attack developed in Ref.~\cite{gluinoA}  for periodic
zero-modes can be carried over if the whole
problem is seen as living in this duplicated space-time. This {\em replica
trick} has been used by some of the authors in other
works~\cite{TwistedNahm,replicas}
and is an important source of information when dealing with periodic
gauge fields. In our case, the problem becomes that of finding
self-dual deformations of the $Q=2$ caloron obtained by the replica
procedure. Notice that the topological
charge is 2 in this case, so we expect 4 (CP-pairs) of self-dual
deformations. Since the gauge field is periodic in the original period,
they can be split into those which are periodic and those
that are antiperiodic in the original period. The former were studied
in our previous paper and correspond to the ordinary deformations of 
the $Q=1$ caloron. Since there are 2 pairs of those,
which are periodic in the {\em small} torus, we expect to find two
pairs of antiperiodic zero-modes.
Unfortunately, although some particular solutions are known~\cite{falk},
there is no analytic general expression for the $Q=2$
caloron which would reduce the study of deformations to the
differentiation of the general solution with respect to the parameters
of the moduli space. In this paper we will thus follow an alternative strategy.
Incidentally our results  could well prove useful in
achieving the goal of obtaining the most general $Q=2$ caloron solution.

The general formula relating deformations to zero-modes in the adjoint
representation is:
\be
\Psi=\frac{1}{2}\delta A_\mu \gamma_\mu (\mathbf{I}\pm \gamma_5) V\,,
\ee 
with the $+$ or $-$ sign depending on whether the solution is self-dual
or antiself-dual. $V$ is an arbitrary constant spinor and hence, 
the zero-modes that we are looking for can be arranged
into two-dimensional complex vector spaces. These spaces are  generated 
by any solution $\Psi$ and its euclidean CP transform
\be
\label{CP_transform}
\Psi \longrightarrow \Psi^c  \equiv \gamma_5 C  \Psi^*\,.
\ee
Our formula can easily be shown to satisfy the Dirac equation provided
the deformation satisfies  the background Lorentz gauge condition.
 \be 
 D_\mu \delta A_\mu=0\,.
 \ee

Using the general ADHM construction one can obtain formulas for the
self-dual deformations in terms of those for the Nahm-ADHM data. The
ADHM construction for SU(2) tells us that a self-dual gauge field can
be constructed as~\cite{adhm}
\begin{equation}
A_\mu(x)= \frac{i}{F}\, (u^\dagger \partial_\mu u)'\,,
\end{equation}
where $u$ is a vector in quaternions, $F=1+u^\dagger u$, and the prime denotes the
traceless part. The vector $u$ is obtained as the solution of the
following equation
\begin{equation}
\label{u_Eq}
(\widetilde{A}^\dagger -x_\mu\overline{\sigma}_\mu) u= q \,,
\end{equation}
where the quaternionic matrix $\widetilde{A}^\dagger$ and the vector
$q$ are  x-independent. We have introduced the Weyl matrix
$\overline{\sigma}_\mu=(\mathbf{I}, i\vec{\tau})$ whose adjoints are
$\sigma_\mu$ ($\vec{\tau}$ are the Pauli matrices).
In the proof of self-duality one must demand
that the following matrix
\begin{equation}
\label{R_Eq}
R\equiv (\widetilde{A}^\dagger -x_\mu\overline{\sigma}_\mu)
(\widetilde{A} -x_\mu\sigma_\mu) + q\otimes q^\dagger\,,
\end{equation} 
is real and invertible. For our purpose  it is interesting to write
down the expression of the adjoint zero-modes in terms of the
deformations of the ADHM data
\begin{equation}
\label{Adj_modes}
\delta A_\mu= \frac{-i}{2} (\delta q^\dagger -u^\dagger \delta
\widetilde{A})\overline{\sigma}^\mu \sigma^\nu \partial_\nu \omega + {\rm h. c.}\,,
\end{equation}
where $\omega =R^{-1} q$. To guarantee that the deformations $\delta q$ 
and $\delta \widetilde{A}$ provide a self-dual deformation $\delta
A_\mu$ satisfying the background field gauge condition, one must impose
certain conditions. These are best expressed in terms of the matrix
with quaternionic entries  ${\cal F} \equiv M^\dagger \delta
M\equiv {\cal F}_\mu \sigma_\mu$, where $M^\dagger=(q, 
\widetilde{A}^\dagger -\bar{\sigma}_\mu x_\mu)$. The condition then reduces to the
hermiticity of ${\cal F}_\mu$ (${\cal F}_\mu={\cal F}_\mu^\dagger$).

The previous  formulas apply for $Q=1$  calorons by extending the vector $q$
to become a delta-like functional over the periodic functions in one-variable
$z$, while $2 \pi i \widetilde{A}$ is a covariant Weyl 
operator with respect to a 1-dimensional abelian gauge field
$\hat{A}_\mu(z)$, the Nahm-dual gauge field. After suitable 
rotations, translations and gauge transformations the caloron Nahm
data  can be taken to be~\cite{vanbaal0,vanbaal}
\be
q^{(0)}(z)= \rho (P_+ \, \delta(z-\delta_1) + P_- \,
\delta(z+\delta_1)) \,,
\ee
where $P_{\pm}=(1 \pm \tau_3)/2$. The 
parameter $\delta_1$ parametrizes the holonomy, becoming trivial
for $0$ and $\frac{1}{2}$. Without
loss of generality we will assume in what follows that $\delta_1
\le \delta_2\equiv\frac{1}{2}-\delta_1$.  
In the previous formula  the delta functions have to be taken as periodic
functions in $z$ with unit period.
The Nahm-dual gauge field of the caloron is given by
\be
\hat{A}^{(0)}_\mu(z)=-2 \pi \delta_{\mu 3} (X_3^1 \chi_1(z)
+X_3^2 \chi_2(z))  \,,
\ee
where $X^a_3$ is the position of the ath constituent monopole on the z-axis.
They can be obtained from the relations
$m_1X_3^1+m_2X_3^2=0$, and $X_3^2-X_3^1=\pi \rho^2$, where $m_a=4\pi\delta_a$
are proportional to the constituent monopole masses.
The function $\chi_1$ is the  characteristic function of the interval
$[-\delta_1,\delta_1]$ and $\chi_2$  that of its complementary.

 Eq.~(\ref{R_Eq}) implies that the Nahm-dual  gauge field is self-dual at all
but a finite number of points. Eq.~(\ref{u_Eq}) is then the solution of the Weyl
equation except at those isolated points. As we will see later the
conditions on the deformations  $\delta \widetilde{A}$  that enter
Eq.~(\ref{Adj_modes}), are precisely equivalent to requiring that
$\delta \hat{A}_\mu$ is again a self-dual deformation satisfying the
background gauge condition. Thus, they can be obtained as the solution
of the  adjoint Weyl equation of the Nahm-dual field, up to delta functions.

Now we should apply this scheme to the replicated caloron taken as a
self-dual solution in the double torus with period $2 \beta$ (remember $\beta$ is fixed to 1).
Since this caloron now has charge $Q=2$
its corresponding Nahm-dual gauge field is now a matrix.
 Using the general construction of Nahm-dual replicas
 given in \cite{TwistedNahm}
we obtain
\be
\hat{A}^{R}_\mu(z)=\pmatrix{\hat{A}^{(0)}_\mu (z) & 0 \cr
0 & \hat{A}^{(0)}_\mu (z+\frac{1}{2} )} \,,
\ee
where $\hat{A}^{(0)}_\mu(z)$ is the Nahm data of the ordinary 
caloron, and $\hat{A}^{R}_\mu(z)$ is the  Nahm data of the replicated
caloron.

One may now wonder which is the corresponding  $q$ for such a replica
solution. We will argue that the solution is actually given by 
\be
q^R(z)=\pmatrix{q^{(0)}(z)\cr q^{(0)}(z+\frac{1}{2})} \,,
\ee
Notice that each of the components of $q$ and $\hat{A}$ are periodic with
unit period, but the whole set is periodic with  period  $1/2$ 
with a twist matrix given by $\tau_1$: 
\be
\hat{A}^R_\mu (z+\frac{1}{2})=\tau_1 \hat{A}^R_\mu(z) \tau_1\,. 
\ee
The quantity $q$ transforms by periodicity as follows:
\be
q^R(z+1/2)= \tau_1 q^R(z)\,.
\ee
From here it is possible to use the general formulas of the ADHM
construction to verify that indeed we obtain a replicated solution.
In particular we have that $u^R(z)$ is given by:
\be
u^R(z)=\pmatrix{u^{(0)}(z)\cr u^{(0)}(z+\frac{1}{2})}\,.
\ee
Now
\be
F^R-1=\int_0^{\frac{1}{2}} dz u^{R \dagger}(z) u^R(z) = \int_0^1 dz\,  u^{(0)
\dagger}(z) u^{(0)}(z) \,,
\ee 
which coincides with $F-1$ for the caloron.  
The replicated gauge potential follows from
\be
A_\mu^R(x)= \frac{i}{ F^R} \int_0^{\frac{1}{2}}  dz (u^{R \dagger}(z)
\partial_\mu u^R(z))' = A_\mu^{(0)}(x)   \,.
\ee

One might  wonder whether the choice of $q$ and $\hat{A}^R_\mu$ are
consistent with the condition of self-duality in Nahm-dual space, namely
that $R=M^\dagger M$ commutes with the quaternions. To verify that this is so,
one must realize that the condition of self-duality should hold only in the
domain of these operators. These are two-component vectors $\psi(z)$
satisfying 
\be
\psi(z+\frac{1}{2})=\tau_1 \psi(z)\,.
\ee
Thus, they should be of the form
\be
\label{Conds}
\psi(z)=\pmatrix{\phi(z)\cr \phi(z+\frac{1}{2})}\,.
\ee
Thus, $q q^\dagger$ acting on this vector yields:
\bea
%\nonumber \\
&& 2 \rho \pmatrix{q^{(0)}(z)\cr
q^{(0)}(z+\frac{1}{2})} (P_+
\phi(\delta_1) +P_- \phi(-\delta_1) )= \nonumber \\ 
&& 2 \rho^2 \pmatrix{P_+
\phi(\delta_1)
\delta(z-\delta_1) +
P_- \phi(-\delta_1)
\delta(z+\delta_1) \cr P_+ \phi(\delta_1)
\delta(z+\delta_2) +
P_- \phi(-\delta_1)
\delta(z-\delta_2) }\,.
\eea
The imaginary part of the upper component coincides with
\be
\rho^2\tau_3\, (\delta(z-\delta_1)-\delta(z+\delta_1))\, \phi(z) \,,
\ee
which is what is needed to cancel the self-duality violation.

Now we proceed  to study the self-dual deformations of this
replicated caloron satisfying the background field condition.
We will make use of our general formula Eq.~(\ref{Adj_modes}).
The conditions following that equation when translated to our case
become
\begin{eqnarray}
\label{Main_eq}
\hat{\bar{D}}\psi^R \equiv \frac{d\psi^R }{dz}-
i\bar{\sigma}_\mu[\hat{A}^R_\mu, \psi^R ] =  4\pi^2i\left(q_\mu^R\delta q_\nu^{\dagger R} -
\delta q_\nu^R q_\mu^{\dagger R}\right)\bar{\sigma}_\mu \sigma_\nu\,,
\end{eqnarray}
where $\psi^R=\delta \hat{A}^R_\mu \sigma_\mu=-\delta \widetilde{A}^R/(2
\pi)$. The quantities $\delta q^R_\nu$ are two-component column vector
whose elements are  linear combinations of delta functions with
complex coefficients.  The holonomy fixes that the argument of the
delta functions must be  $z\pm \delta_1$ and $z\pm
\delta_1+\frac{1}{2}$. Notice that, as anticipated previously, up to
the delta functions in the right-hand side, the equation adopts the form of the Weyl
equation for adjoint zero-modes in Nahm dual space.

Our next step will then be that of finding the solution of
Eq.~(\ref{Main_eq}). Notice that both $\psi^R$ and $\delta q^R\equiv
\delta q^R_\nu \bar{\sigma}_\nu$ are the unknowns. Without much effort
one can demonstrate that given a solution one can obtain other
solutions by the operation $\psi^R \rightarrow \psi^R  Q$,
$\delta q^R\rightarrow Q^\dagger \delta q^R$, with $Q$ an arbitrary
constant quaternion. This transformation is associated to the double
degeneracy of adjoint zero-modes. We must also point out certain
subtleties necessary to understand  Eq.~(\ref{Main_eq}) and their solutions.
The main idea is that  the equation must be understood as one  relating
two operators  acting on the space two-component  functions of the form
Eq.~(\ref{Conds}). The right-hand side of Eq.~(\ref{Main_eq}) acts by
multiplication. Thus, the left-hand side must be equivalent,
when acting over our space of functions, to the multiplication by a
linear combination of delta functions. This imposes non-trivial
conditions on the form of $\delta q^R$. In what follows we will give the
possible values for $\delta q^R$ that follow from the previous analysis,
as well as the resulting form for the equation for $\psi^R$, skipping
all the details of the derivation.

Before showing the equations, we recall that $\psi^R$ is a $2\times 2 $
matrix in (Nahm-dual) colour space
\be
\psi^R (z)=\pmatrix{\psi_{11}(z) & \psi_{12}(z) \cr \psi_{21}(z) & \psi_{22}(z)}\,.
 \ee
The boundary conditions specify that it is enough to know the form
of $\psi_{11}$ and $\psi_{12}$ (the other components can be obtained
by translating in $z$ by $1/2$). The equations for $\psi_{11}$
coincide with those for the $Q=1$ caloron, and therefore can be
associated with deformations that are periodic in time. Thus, our
sought time-antiperiodic zero-modes should follow from the equation
\bea
\label{psi_eq}
\partial_z \psi_{12} + \tau_3 (\Delta \hat{A}) \psi_{12} =  
4 \pi^2 \rho \Big \{\, P_+ &\Big(&\delta(z-\delta_1)-\delta(z-\delta_2)\Big)\nonumber \\
+ P_- &\Big(&\delta(z+\delta_2)-\delta(z+\delta_1)\, \Big) \Big\} \, Q\,,
\eea
where the function $\Delta \hat{A}$ is given by
\bea
\Delta \hat{A} \equiv \hat{A}^{(0)}_3(z)-\hat{A}^{(0)}_3(z+1/2) = 2\pi^2
\rho^2  (\chi(-\delta_1,\delta_1) -  
\chi(\delta_2,1-\delta_2))\,.
\eea
The arbitrary quaternion $Q$ reflects the degeneracy of solutions
mentioned earlier. Keeping that in mind one  only needs to solve
the equation for $Q=0$ and $Q=1$. A particular solution is all that is
needed, since the general solution can be obtained by linear
combinations of these ones with quaternionic coefficients.
The counting matches the predictions of the index theorem. As for the
periodic case there are essentially two CP-pairs  of zero-modes. 
 
After these considerations we proceed to show the two particular
solutions that we will need. The first one corresponds to the
inhomogeneous equation ($Q=1$) and is given by
\be
\psi_{1 2}= 4 \pi^2 \rho \left(P_+\chi(\delta_1, \delta_2) +
P_-\chi( 1-\delta_2,1-\delta_1)\right)\,.
\ee
The value of $\delta q^R$ associated to it is
\be
\delta q^R=i P_+ \pmatrix{\delta(z+\delta_2) \cr
\delta(z-\delta_1)} - i P_-
\pmatrix{\delta(z-\delta_2) \cr
\delta(z+\delta_1)}
\ee
These expressions  can now be introduced into the general formula Eq.~(\ref{Adj_modes}) to obtain the first solution
\bea
\nonumber
&&\delta A_\mu^{(1)}= -\frac{1}{2}\Big(P_+ \bar{\sigma}_\mu \hat{\partial}
\omega(-\delta_2) - P_- \bar{\sigma}_\mu \hat{\partial}
\omega(\delta_2)\Big)\\&&
\label{EQONE}
- i \pi \rho \Big( \int_{\delta_1}^{\delta_2} u^\dagger(z+\frac{1}{2})
P_-\bar{\sigma}_\mu \hat{\partial} \omega(z) + \int_{1-\delta_2}^{1-\delta_1} u^\dagger(z+\frac{1}{2})
P_+\bar{\sigma}_\mu \hat{\partial} \omega(z) \Big) + {\rm h. c.}\,.
\eea
The quantities $u$ and $\omega$ are the ones associated to the $Q=1$
caloron. The analytic expressions needed to do the calculation were
explicitly given in our previous paper~\cite{gluinoA}.

Now we investigate the other solution, associated to $\delta q^R=0$.
One has to solve the homogeneous equation~(\ref{psi_eq}) for vanishing
right hand side.  A particular solution is given  by
\bea
\psi_{ 1 2}(z) =  \exp\{ - \tau_3 \int_0^z dz' \, \Delta \hat{A}(z')\}
\equiv \phi_s(z) -\tau_3 \phi_a(z) \,.
\eea
Since $\Delta \hat{A}(z')$ is constant at intervals, the integral in the exponent is trivial to perform. We leave the explicit form of
$\phi_s(z)$ and $\phi_a(z)$ to the reader. It is interesting to point
out nonetheless, that $\phi_s(z)$ is periodic in $z$ with period
$\frac{1}{2}$ and $\phi_a(z)$ antiperiodic.

From the previous expression  we can compute the corresponding self-dual
deformation using Eq.~(\ref{Adj_modes}). The result is given by
\bea
\label{EQTWO}
\delta A^{(2)}_\mu = \frac{-i}{ 4 \pi} \int_0^1 dz \
(u^\dagger(z+\frac{1}{2})
(\phi_s(z) + \tau_3 \phi_a(z)) \bar{\sigma}_\mu \hat{\partial}\omega(z))
+ \hc \,.
\eea
Again, the integration over $z$ can be performed analytically using the
formulas of our previous paper~\cite{gluinoA}.

We have arrived to the general solution  our problem.
The adjoint zero-modes of the (self-dual) caloron which are
antiperiodic in time are
\be
\label{eq.gen}
\Psi=\frac{1}{2}\delta A^{(1)}_\mu \gamma_\mu (\mathbf{I}+ \gamma_5)
V_1 + \frac{1}{2}\delta A^{(2)}_\mu \gamma_\mu (\mathbf{I}+ \gamma_5)
V_2\,,
\ee
where $V_a$ are arbitrary constant spinors and $\delta A^{(a)}_\mu$
are given in Eqs.(\ref{EQONE})-(\ref{EQTWO}).

It is interesting to mention that the general investigation of
the possible values of $\delta q^R$ has led to another solution having
a fairly simple form. The expression of the left-handed Weyl spinor,
$\Psi^{(3)}\equiv \Psi_a^{(3)} \tau_a$, is:
\be
\label{ADDSOL}
\Psi_a^{(3)}= \sigma_\mu \partial_\mu T^a \sigma_a   V\,,
\ee
where $a$ labels a colour component, $\sigma_\alpha$ acts on the spin indices
and $V$ denotes an arbitrary constant 2-spinor.
The functions $T_a$ depend on the colour index as: $T^1=T^2=-1/F$ and
$T^3= P_+ \chi + P_- \bar{\chi}$. The function $\chi$ is essentially the
function with the same name given in
Ref.~\cite{vanbaal0,vanbaal}.
Curiously this solution interpolates between the
non-supersymmetric  periodic adjoint zero-mode for $m_1=0$ ($\delta_1=0$)
and one of our  antiperiodic solutions (Eq.~(\ref{EQONE})) for
$m_1=m_2$ ($\delta_1=1/4$). Using the formulas of the next section
it can be proven that the solution is neither periodic non antiperiodic for other values
of the mass $m_1$.

\section{Properties of the solutions}
\label{s.prop}
In this section we will investigate the general properties of the
solutions found in the previous section.

\subsection{Periodicity in time}

Here  we will explicitly verify the required antiperiodicity in time
of our general solution. In our gauge the caloron vector potential
satisfies
\be
A_\mu^{(0)}(x_0+1)= e^{i {m_1 \tau_3\over 2}} A_\mu^{(0)}(x_0)e^{-i {m_1 \tau_3\over 2}}
\ee
Thus, the required antiperiodicity of the adjoint zero-modes amounts to:
\be
\Psi(x_0+1)= -e^{i {m_1 \tau_3\over 2}} \Psi(x_0)e^{-i {m_1 \tau_3\over 2}}
\ee
This property follows easily from the form of our solutions
and the periodicity behaviour of $u$:
\be
u(z,x_0+1)=e^{i2 \pi z} u(z,x_0) e^{-i {m_1 \tau_3\over 2}}
\ee
and an identical  relation  for $\omega$ and $q$.

\subsection{Far-field limit and Normalization}
The reader might question whether our general solution Eq.~(\ref{eq.gen})
is normalizable. One can investigate the behaviour at points whose
distance to the location of the constituent monopoles  ($r_1$ and
$r_2$) is much larger that $\beta$ and that $\pi \rho^2$. For the unequal
mass case the zero-mode density goes to zero exponentially as
$e^{-(m_2-m_1)r_2}$. The equal mass case ($m_1=m_2=\pi$) is more subtle
since both solutions decay in power-like fashion.
The non-homogeneous solution Eq.~(\ref{EQONE}) coincides with the
additional solution Eq.~(\ref{ADDSOL}) in this case. In the limit under
consideration $\chi$ goes to zero exponentially and $F=(r_1+r_2+\pi
\rho^2)/(r_1+r_2-\pi \rho^2) $. Thus the density behaves as $1/r^4$.

An alternative approach to normalizability of the solutions is to
compute the norm of the solutions. In fact there exist a   general
formula~\cite{vanbaal0,kraan} which allows one to compute the norm and the scalar
products of the solutions in terms of Nahm-data directly. This is
also useful in checking if the real dimensionality of the space of
solutions is 8 (4 complex dimensions, 2 quaternionic dimensions),
as indicated by the index theorem. Using this formula we obtain
\begin{eqnarray}
&& \left|\delta A_\mu^{(1)}\right|^2 = 4\pi^2+8\pi^4\rho^2\,(\delta_2-\delta_1)\,, \\
&& \left|\delta A_\mu^{(2)}\right|^2 = \frac{\sinh(4\pi^2\rho^2\delta_1)}{2\pi^2\rho^2}+(\delta_2-\delta_1)\,\cosh(4\pi^2\rho^2\delta_1)\,, \\ 
&& \left<\delta A_\mu^{(1)},\delta A_\mu^{(2)}\right> = 2\pi^2\rho \,\e^{-2\pi^2\rho^2\delta_1}\,(\delta_2-\delta_1)\,.
\end{eqnarray}

\subsection{Profile  of the zero-mode density}
\begin{figure}
 \centerline{
\psfig{file=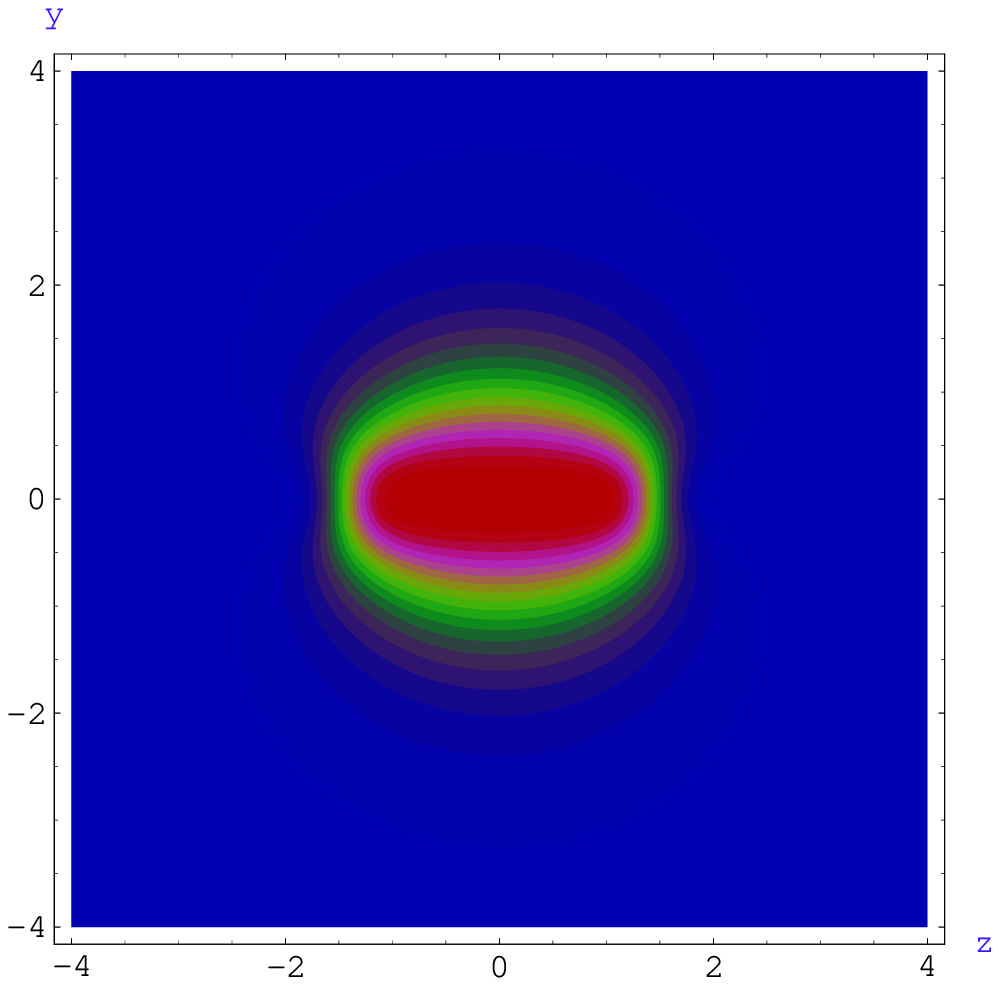,width = 6 cm}
\hspace{0.3cm} \psfig{file=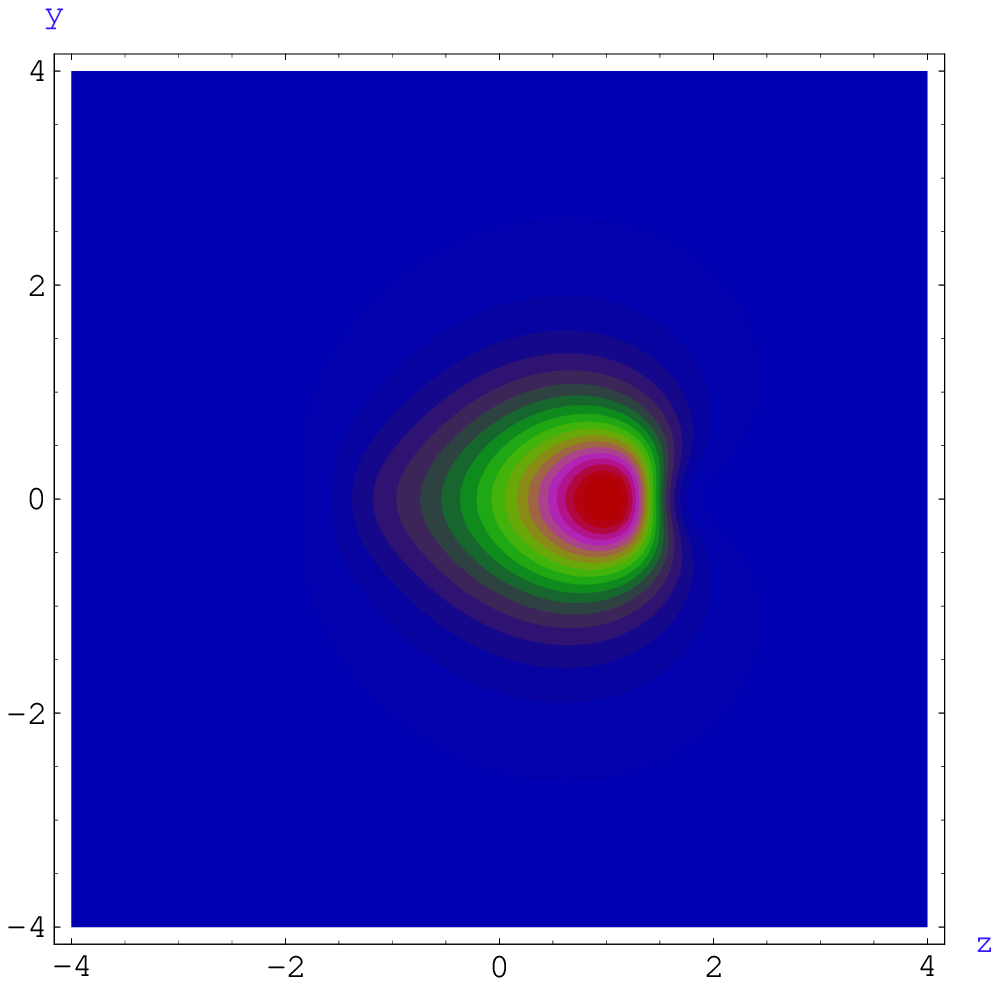,width = 6 cm}
}
 \centerline{
\psfig{file=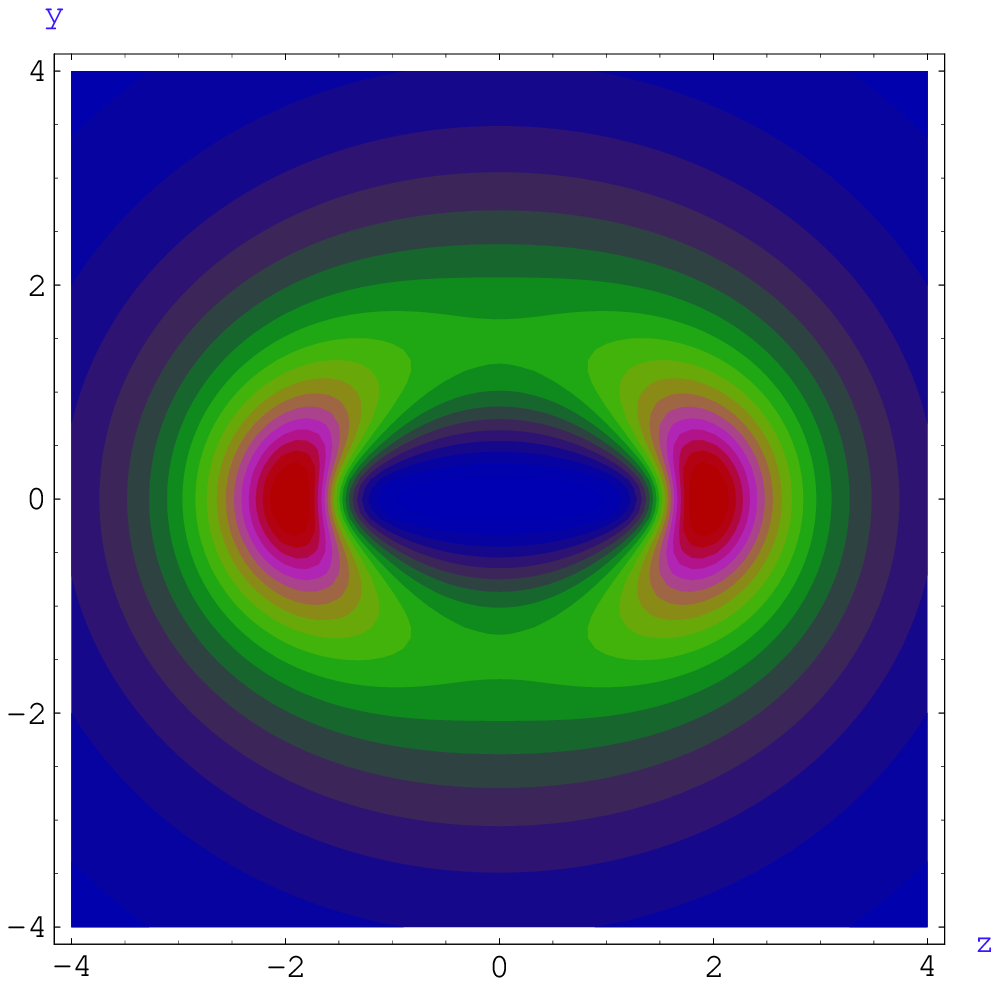,width = 6 cm}
\hspace{0.3cm} \psfig{file=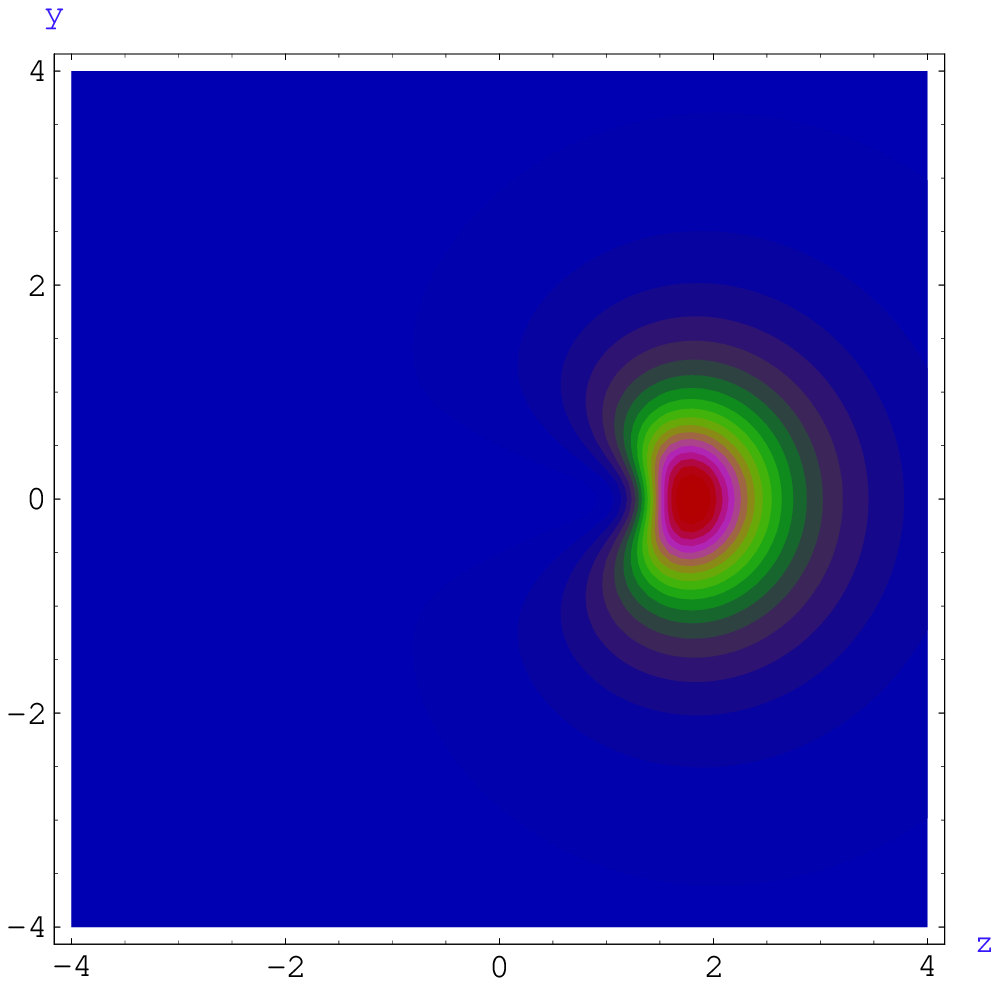,width = 6 cm}}
\centerline{\psfig{file=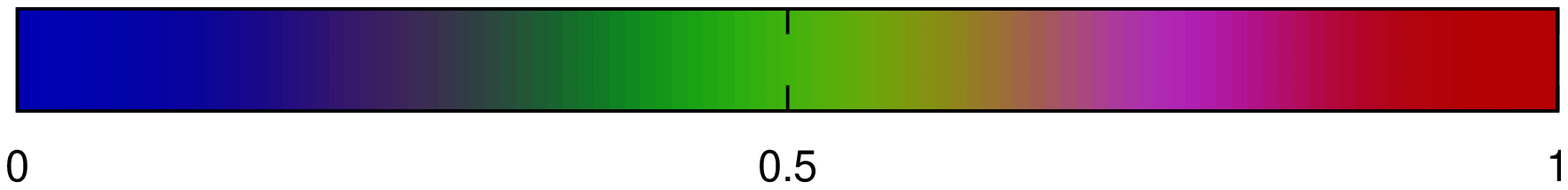,width = 1 cm,angle = 270}
}

\caption{Contour plots of the density of the two antiperiodic zero modes
in the $y-z$ plane. Constituent monopoles are localized at $y=0$ and separated
along the z axis which is drawn horizontally. Left: For $m_1=m_2=\pi$ and $\rho=1$.
Right: For $\delta_1=0.23$ and $\rho=1$.}
\label{fig:fig1}
\end{figure}

In this subsection we will describe the qualitative properties
of the zero-mode densities. For that purpose we developed two
independent programs to draw these profiles. Both programs give 
matching results. In Fig.~\ref{fig:fig1} we give the contour plot in a z-y plane
of the solution $\delta A^{(2)}_\mu$ (top) and an orthogonal CP-pair (bottom) for $\rho=1$ (giving an intermediate size caloron
separation) and  two representative values of the masses.
The $z$ axis 
is the line joining the constituent monopoles and is represented 
horizontally. The vertical axis denotes the $y$ axis (the density  
is axially symmetric).

For the equal mass case ($m_1=m_2=\pi$) the mode following from Eq.~(\ref{EQTWO})
has an approximately constant higher density along the line joining
both calorons (top left). This can be interpreted as a string.
In contrast,  the other solution associated
 to Eq.~(\ref{EQONE}) has a region of small density located
 along the line joining the two monopoles (bottom left). As the masses
 become unequal, the most massive monopole dominates the densities.
 The right contour plots show the situation for $\delta_1=0.23$.

\subsection{Limiting cases}
The caloron is an interesting solution which interpolates between
the gauge potential of an instanton and that of a BPS monopole. It is
interesting then to see how our antiperiodic zero-modes behave in these
extreme cases. We will first concentrate in the situation corresponding to 
the trivial
holonomy, Harrington-Shepard, caloron: $\delta_1=0$. In that case one of the
constituent monopoles is massless and pushed to infinity. The $\rho$ parameter
of the solution does no longer control the separation between the monopoles
but is still a free parameter. For small $\rho$ the HS caloron approaches an
ordinary, zero temperature, instanton. From our general formulas, it is easy to check
that in that limit and close to the center of the caloron the
time periodicity becomes irrelevant and the two zero mode CP-pairs approach the
periodic zero modes of the instanton. In the opposite, $\rho\rightarrow \infty$, limit
the HS caloron becomes a BPS monopole with time independent action density.
Despite the time independence of the background there are still 4 non-trivial
antiperiodic zero-modes. They can be easily derived from Eqs. ~(\ref{EQONE})
and (\ref{EQTWO}) by taking the appropriate $\delta_1=0$ and $\rho\rightarrow \infty$
limits. Up to a gauge transformation we obtain:
\bea
\delta A^{(1)'}_\mu(x)&=&  \eta_{3\mu}^\alpha \, \pi \rho\, \Big(\bar e_1^2(x) \,  E_\alpha^{\rm bps}(r)  
- \bar e_2^2(x)  \,\tilde E_\alpha(r) \Big) + \hc \,,\\ 
\delta A^{(2)}_\mu(x) &=& {1 \over 4\pi} \, \Big(\bar e_2^2(x)\,  E_\mu^{\rm bps}(r) + \bar e_1^2(x)\, \tilde E_\mu(r) \Big)
+ \hc\,,
\eea
where $\delta A^{(2)}_\mu $ is directly derived from Eq.~(\ref{EQTWO}) and
$\delta A^{(1)'}_\mu$ is the combination of  
Eqs.~(\ref{EQONE}) and (\ref{EQTWO}) orthogonal to $\delta A^{(2)}_\mu$.
In the expression above, $E_\alpha^{\rm bps}$ is the electric
field of the BPS monopole:
\be
E_\alpha^{\rm bps}(x) = -i\, {g^2(2\pi r) -1 \over 2 r^2} \, P_\alpha^+ -
i\, {\pi g'(2\pi r)  \over r} \, P_\alpha^-\,,
\ee
and we have introduced the time independent quantity:
\be
\tilde E_\alpha (x) = {\tanh(\pi r)\over 2 \cosh(\pi r)} \, \Big (i\, {g(\pi r) -\cosh(\pi r)\over  r^2} \,  P_\alpha^+
-i \,{\pi g'(\pi r)  \over  r } \, P_\alpha^-\Big )\,,
\ee
with $g(u) = u/\sinh(u)$, $g'(u)$ its derivative with respect to $u$,  and
$P_\mu^\pm = (\bar \sigma_\mu \pm \hat n \bar \sigma_\mu \hat n)/2$,
$\hat n = x_i \tau_i/r$. The antiperiodicity of the solution is encoded
in the time dependent quaternions $\bar e_1^2$ and $\bar e_2^2$ defined through:
\be
e^{-i \pi x_\mu \bar \sigma_\mu} = i (e_1^2(x)+ i e_2^2(x))\,.
\ee

For non-trivial holonomy there is also an interesting limit in
which the caloron solution tends to the BPS monopole.
It corresponds to making the separation of the constituent monopoles 
tend to infinity ($\rho \rightarrow \infty$). Our adjoint 
zero-modes lead to those of the BPS monopole if the appropriate limit
is taken ($r_1<< \pi \rho^2$, $\rho >> 1$). 
For example, for the equal mass case  ($m_1=m_2=\pi$) the first solution
Eq.~(\ref{EQONE}) follows quite simply by applying the appropriate limit to
Eq.~(\ref{ADDSOL}). Computing the density we obtain:
\begin{equation}
2 (h'^2(\pi r) +1 -2 h'(\pi r)\cos\theta)+ g^2(\pi r) + g'^2(\pi r)+
2 g(\pi r)  g'(\pi r)\cos\theta 
\end{equation}
where $h'(u)$ is the derivative of $h(u)\equiv u\coth(u)$. 
This profile has axial symmetry depending explicitly on the azimuthal
angle $\theta$.  Notice also that the  solution is non-normalizable.

\subsection{Comparison with numerical results}
\begin{figure}
 \centerline{
\psfig{file=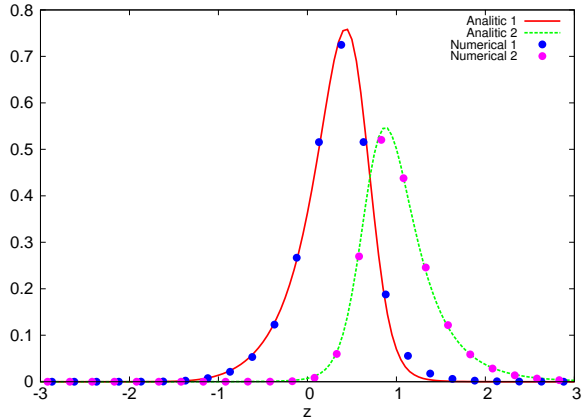,angle=-90, width = 8 cm}
}
\caption{Comparison between numerical (circles) and analytic (lines) zero modes
for $\rho=0.79$ and $\delta_1=0.172$.
We display the density of the zero modes, integrated in time, along the line
joining the two monopoles.}
\label{fig:fig2}
\end{figure}

We have crosschecked our results with a direct evaluation of adjoint
zero-modes on the torus obtained by lattice methods using Neuberger's
overlap operator~\cite{neuberger} in the adjoint representation. One
expects that the spatial profile of the torus solutions approaches our
analytical formulas as the box size becomes much larger  than all scales
of the problem ( $\beta$ and $\pi \rho^2$). To make a quantitative comparison
we computed  the zero-mode density integrated in time along the line
$x=y=0$ joining both constituent monopoles. It is not possible a
priori to construct numerical zero-modes with a prescribed value of
$\rho$ and $\delta_1$, although some tuning is possible~\cite{cal_lat}.
For the numerical comparison displayed in Fig.~\ref{fig:fig2} we slightly
tuned by hand these parameters to improve the agreement ($\rho=0.79 $,
$\delta_1=0.172 $). A technical point which one has to address is
how to guarantee that the same linear combinations are selected for the
numerical and analytical data. We chose to define the two linearly
independent  modes  by imposing that at the center of mass
($x=y=z=0$) one has maximal and the other minimal density (integrated  
over time). 

\section{Conclusions}
\label{s.concl}
In this paper we have obtained analytic formulas for the zero-modes
of the Dirac equation for gluinos in the background field of Q=1
SU(2) calorons with antiperiodic boundary conditions in thermal-time.
Our formulas are valid for non-trivial holonomy as well
as for the Harrington-Shepard caloron and include as a limiting case
those of BPS monopoles. The solutions have finite norm and decay
exponentially with distance if the masses of the constituent monopoles
differ. Their density profile contrasts with the case of periodic
zero-modes. For example, as the monopoles are pulled apart the density
does not decouple into independent lumps centered at the monopoles,
but rather describes a string  joining the monopoles. Nonetheless, the
number of normalizable zero-modes matches in both cases.

Our work has  methodological interest since  our
approach  is applicable to other cases including the extension to
SU(N), and might be instrumental in finding formulas for calorons of
higher charge. From a physical viewpoint our work provides a first step
towards a semiclassical study of  N=1 SUSY Yang-Mills at finite temperature.
There are interesting  issues at stake such as  that  of supersymmetry
breaking at finite temperature, which  has been a subject of debate since
early  times~\cite{Das:1978rx}-\cite{Buchholz:1997mf}. It is our intention to
address these questions in  future work.

\section*{Acknowledgments}
We would like to thank Falk Bruckmann for discussions at the initial
stages of this work.
We acknowledge financial support from  Comunidad Aut\'onoma
de Madrid under the program  HEPHACOS P-ESP-00346. A.S. is supported by an FPU
fellowship of Spanish Research Ministry (MEC).
M.G.P. and A.G-A acknowledge financial support from grants from CICYT
FPA2006-05807, FPA2006-05485 and FPA2006-05423 . The authors participate in the Consolider-Ingenio 2010 CPAN (CSD2007-00042). 
We acknowledge the use of the IFT cluster for part of our numerical results.

% \appendix

% \section{}
% \label{}


\begin{thebibliography}{00}
\bibitem{HS}
B.~J.~Harrington and H.~K.~Shepard,
  %``Periodic Euclidean Solutions And The Finite Temperature Yang-Mills Gas,''
  Phys.\ Rev.\ D {\bf 17}, 2122 (1978).
  %%CITATION = PHRVA,D17,2122;%%
%``Thermodynamics Of The Yang-Mills Gas,''
  Phys.\ Rev.\ D {\bf 18} (1978) 2990.
  %%CITATION = PHRVA,D18,2990;%%


\bibitem{vanbaal0}
 T.~C.~Kraan and P.~van Baal,
%``Periodic instantons with non-trivial holonomy,''
  Nucl.\ Phys.\ B {\bf 533} (1998) 627
  [arXiv:hep-th/9805168].
  %%CITATION = HEP-TH 9805168;%%

\bibitem{vanbaal}
 T.~C.~Kraan and P.~van Baal,
 %``Exact T-duality between calorons and Taub - NUT spaces,''
 Phys.\ Lett.\ B {\bf 428} (1998) 268
 [arXiv:hep-th/9802049].
 %%CITATION = HEP-TH 9802049;%%
 T.~C.~Kraan and P.~van Baal,
  %``Monopole constituents inside SU(n) calorons,''
  Phys.\ Lett.\ B {\bf 435} (1998) 389
  [arXiv:hep-th/9806034].
  %%CITATION = HEP-TH 9806034;%%

\bibitem{lee}
K.~M.~Lee,
  %``Instantons and magnetic monopoles on R**3 x S(1) with arbitrary simple
  %gauge groups,''
  Phys.\ Lett.\ B {\bf 426} (1998) 323
  [arXiv:hep-th/9802012].
  %%CITATION = HEP-TH 9802012;%%
K.~M.~Lee and C.~h.~Lu,
  %``SU(2) calorons and magnetic monopoles,''
  Phys.\ Rev.\ D {\bf 58}, 025011 (1998)
  [arXiv:hep-th/9802108].
  %%CITATION = HEP-TH 9802108;%%

\bibitem{Rossi}
 P.~Rossi,
  %``Exact Results In The Theory Of Nonabelian Magnetic Monopoles,''
  Phys.\ Rept.\  {\bf 86} (1982) 317.
  %%CITATION = PRPLC,86,317;%%

\bibitem{grossman}
B. Grossman,
  Phys.\ Lett.\ A {\bf 61} (1977) 86

\bibitem{Bilic}
  N.~Bilic,
  %``Fermion Zero Modes In Finite Temperature Yang-Mills Theory,''
  Phys.\ Lett.\  B {\bf 97} (1980) 107.
  %%CITATION = PHLTA,B97,107;%%

\bibitem{Simonov}
  A.~Gonzalez-Arroyo and Yu.~A.~Simonov,
  %``Fermionic zero modes for dyons and chiral symmetry breaking in QCD,''
  Nucl.\ Phys.\  B {\bf 460} (1996) 429
  [arXiv:hep-th/9506032].
  %%CITATION = NUPHA,B460,429;%%

\bibitem{us}
  M.~Garc\'{\i}a P\'erez, A.~Gonz\'alez-Arroyo, C.~Pena and P.~van Baal,
  %``Weyl-Dirac zero-mode for calorons,''
  Phys.\ Rev.\ D {\bf 60} (1999) 031901
  [arXiv:hep-th/9905016].
  %%CITATION = HEP-TH 9905016;%%

\bibitem{maxim}
 M.~N.~Chernodub, T.~C.~Kraan and P.~van Baal,
  %``Exact fermion zero-mode for the new calorons,''
  Nucl.\ Phys.\ Proc.\ Suppl.\  {\bf 83} (2000) 556
  [arXiv:hep-lat/9907001].
  %%CITATION = HEP-LAT 9907001;%%
  F.~Bruckmann, D.~Nogradi and P.~van Baal,
  %``Constituent monopoles through the eyes of fermion zero-modes,''
  Nucl.\ Phys.\ B {\bf 666} (2003) 197
  [arXiv:hep-th/0305063].
  %%CI

  \bibitem{gluinoA}
 M.~Garc\'{\i}a P\'erez and A.~Gonz\'alez-Arroyo,
  %``Gluino zero-modes for non-trivial holonomy calorons,''
  JHEP {\bf 0611} (2006) 091
  [arXiv:hep-th/0609058].
  %%CITATION = JHEPA,0611,091;%%
  
 \bibitem{TwistedNahm}
A.~Gonz\'alez-Arroyo,
  %``On Nahm's transformation with twisted boundary conditions,''
  Nucl.\ Phys.\  B {\bf 548} (1999) 626
  [arXiv:hep-th/9811041]. 
  %%CITATION = NUPHA,B548,626;%%
  
  \bibitem{replicas}
A.~Gonz\'alez-Arroyo and C.~Pena,
  %``Nahm transformation on the lattice,''
  JHEP {\bf 9809} (1998) 013
  [arXiv:hep-th/9807172]. 
  %%CITATION = JHEPA,9809,013;%% 
  M.~Garc\'{\i}a P\'erez, A.~Gonz\'alez-Arroyo, C.~Pena and P.~van Baal,
  %``Nahm dualities on the torus: A synthesis,''
  Nucl.\ Phys.\  B {\bf 564} (2000) 159
  [arXiv:hep-th/9905138].
  %%CITATION = NUPHA,B564,159;%%

 \bibitem{falk}
  F.~Bruckmann and P.~van Baal,
  %``Multi-caloron solutions,''
  Nucl.\ Phys.\ B {\bf 645}, 105 (2002)
  [arXiv:hep-th/0209010].
  %%CITATION = HEP-TH 0209010;%%
 F.~Bruckmann, D.~Nogradi and P.~van Baal,
  %``Higher charge calorons with non-trivial holonomy,''
  Nucl.\ Phys.\ B {\bf 698} (2004) 233
  [arXiv:hep-th/0404210].
  %%CITATION = HEP-TH 0404210;%%

\bibitem{adhm}
  M.~F.~Atiyah, N.~J.~Hitchin, V.~G.~Drinfeld and Y.~I.~Manin,
  %``Construction Of Instantons,''
  Phys.\ Lett.\ A {\bf 65} (1978) 185.
  %%CITATION = PHLTA,A65,185;%%

\bibitem{nahm}
  W.~Nahm,
  %``A Simple Formalism For The Bps Monopole,''
  Phys.\ Lett.\ B {\bf 90} (1980) 413;
  %%CITATION = PHLTA,B90,413;%%
  ``All Selfdual Multi - Monopoles For Arbitrary Gauge Groups,''
CERN-TH-3172
%\href{http://www.slac.stanford.edu/spires/find/hep/www?r=cern-th-3172}{SPIRES entry}
{\it Presented at Int. Summer Inst. on Theoretical Physics, Freiburg, West Germany, Aug 31 - Sep 11, 1981}


\bibitem{neuberger}
 H.~Neuberger,
  %``A practical implementation of the overlap-Dirac operator,''
  Phys.\ Rev.\ Lett.\  {\bf 81} (1998) 4060
  [arXiv:hep-lat/9806025];
  %%CITATION = HEP-LAT 9806025;%%
  %``More about exactly massless quarks on the lattice,''
  Phys.\ Lett.\ B {\bf 427} (1998) 353
  [arXiv:hep-lat/9801031];
  %%CITATION = HEP-LAT 9801031;%%
  %``Exactly massless quarks on the lattice,''
  Phys.\ Lett.\ B {\bf 417} (1998) 141
  [arXiv:hep-lat/9707022].
  %%CITATION = HEP-LAT 9707022;%%
  
  
  \bibitem{cal_lat}
 M.~Garc\'{\i}a P\'erez, A.~Gonz\'alez-Arroyo, A.~Montero and P.~van Baal,
  %``Calorons on the lattice: A new perspective,''
  JHEP {\bf 9906} (1999) 001
  [arXiv:hep-lat/9903022]. 
  %%CITATION = JHEPA,9906,001;%%


\bibitem{kraan}
  T.~C.~Kraan,
  %``Instantons, monopoles and toric hyperKaehler manifolds,''
  Commun.\ Math.\ Phys.\  {\bf 212} (2000) 503
  [arXiv:hep-th/9811179].
  %%CITATION = HEP-TH 9811179;%%

\bibitem{Das:1978rx}  A.~K.~Das and M.~Kaku,
  %``Supersymmetry At High Temperatures,''
  Phys.\ Rev.\  D {\bf 18} (1978) 4540.
  %%CITATION = PHRVA,D18,4540;%%

\bibitem{Girardello:1980vv}
  L.~Girardello, M.~T.~Grisaru and P.~Salomonson,
  %``Temperature And Supersymmetry,''
  Nucl.\ Phys.\  B {\bf 178} (1981) 331.
  %%CITATION = NUPHA,B178,331;%%

\bibitem{VanHove:1982cc}
  L.~Van Hove,
  %``Supersymmetry And Positive Temperature For Simple Systems,''
  Nucl.\ Phys.\  B {\bf 207} (1982) 15.
  %%CITATION = NUPHA,B207,15;%%

\bibitem{Aoyama:1984bk}
  H.~Aoyama and D.~Boyanovsky,
  %``Goldstone Fermions In Supersymmetric Theories At Finite Temperature,''
  Phys.\ Rev.\  D {\bf 30} (1984) 1356.
  %%CITATION = PHRVA,D30,1356;%%

\bibitem{Boyanovsky:1983tu}
  D.~Boyanovsky,
  %``Supersymmetry Breaking At Finite Temperature: The Goldstone Fermion,''
  Phys.\ Rev.\  D {\bf 29} (1984) 743.
  %%CITATION = PHRVA,D29,743;%%

\bibitem{Matsumoto:1984ew}
  H.~Matsumoto, M.~Nakahara, Y.~Nakano and H.~Umezawa,
  %``A New Zero Energy Mode In Supersymmetry At Finite Temperature,''
  Phys.\ Lett.\  B {\bf 140} (1984) 53;
  %%CITATION = PHLTA,B140,53;%%
%``Supersymmetry At Finite Temperature,''
  Phys.\ Rev.\  D {\bf 29} (1984) 2838.
  %%CITATION = PHRVA,D29,2838;%%

\bibitem{Buchholz:1997mf}
  D.~Buchholz and I.~Ojima,
  %``Spontaneous collapse of supersymmetry,''
  Nucl.\ Phys.\  B {\bf 498} (1997) 228
  [arXiv:hep-th/9701005].
  %%CITATION = NUPHA,B498,228;%%
  


\end{thebibliography}
\end{document}